\documentclass[aps,prl,amsmath,amssymb,preprint,superscriptaddress]{revtex4-2} 

\usepackage{graphicx}
\usepackage{dcolumn}
\usepackage{bm}
\usepackage[utf8]{inputenc}
\usepackage[T1]{fontenc}
\usepackage{mathptmx}
\usepackage{etoolbox}
\usepackage{color}
\usepackage{hyperref}
\usepackage{booktabs} 
\usepackage{array}
\usepackage{float}    

\makeatletter
\def\@email#1#2{
 \endgroup
 \patchcmd{\titleblock@produce}
  {\frontmatter@RRAPformat}
  {\frontmatter@RRAPformat{\produce@RRAP{*#1\href{mailto:#2}{#2}}}\frontmatter@RRAPformat}
  {}{}
}
\makeatother

\begin{document}
\begin{sloppypar}

\title{Mechanism of $E'_\gamma$ Defect Generation in Ionizing-irradiated $a$-SiO$_2$: The Nonradiative Carrier Capture-Structural Relaxation Model} 

\author{Yu Song} 
\affiliation{
Xinjiang Key Laboratory of Extreme Environment Electronics, Xinjiang Technical Institute of Physics and Chemistry, Chinese Academy of Sciences, Urumqi 830011, China}
\altaffiliation{Previous at Neijiang Normal University, Neijiang 641112, China}

\author{Chen Qiu}
\affiliation{Eastern Institute of Technology, Ningbo 315200, China}

\author{Su-Huai Wei$^*$}
\email{suhuaiwei@eitech.edu.cn}
\affiliation{Eastern Institute of Technology, Ningbo 315200, China}

\date{May 13, 2025} 

\begin{abstract} 
The total ionizing dose (TID) effect of semiconductor devices stems from radiation-induced $E'_\gamma$ defects in the $a$-SiO$_2$ dielectrics, but the conventional ``hole transport-trapping'' model of defect generation fails to explain recent basic experiments.
Here, we propose an essentially new ``nonradiative carrier capture-structural relaxation'' (NCCSR) mechanism that can consistently explain the puzzling temperature/electric-field dependence, {based on spin-polarized HSE06 hybrid functional calculations and existing experimental alignment of} defect formation energies and charge capture cross-sections of large-sample oxygen vacancies in $a$-SiO$_2$.
It is revealed that, the long-assumed $V_{O\gamma}$ precursors with high formation energy
cannot survive in high temperature-grown $a$-SiO$_2$; whereas the stable $V_{O\delta}$ can capture irradiation-induced holes via strong electron-phonon coupling, 
generating metastable $E'_\delta$ that most relax into stable $E'_\gamma$.
A fractional power-law (FPL) dynamic model is derived based on the mechanism and the Kohlrausch-Williams Watts (KWW) decay function.
It can uniformly describe nonlinear data over a wide dose and temperature range. 
This work not only provides a solid cornerstone for ‌prediction and hardening of TID effects of SiO$_2$-based semiconductor devices, but also offers a general approach for studying ionizing radiation physics in alternative dielectrics with intrinsic electronic metastability and dispersion.
\end{abstract}

\maketitle

{\textit{Introduction.}--}
Due to its easy growth and excellent insulating properties, 
amorphous silicon dioxide ($a$-SiO$_2$) is often used as dielectric or isolation layers in various semiconductor devices, which are made of conventional Si~\cite{prinzie2021low},
emerging two-dimensional (2D) materials~\cite{das2021transistors,akinwande2019graphene},
or wide-band-gap (WBG) SiC~\cite{alkauskas2008band}.
However, these devices suffer from permanent total ionizing dose (TID) damages (e.g., shifts of threshold voltages 
and increases in leakage currents)  
when used in outer space, nuclear reactor, or particle-accelerator applications. 
This is because various defects such as $E'_\gamma$, $E'_\delta$, and $E'_\alpha$ centers  
are induced in the $a$-SiO$_2$ layer by high-energy {ionizing 
irradiations~\cite{oldham2003total,schwank2008radiation,fleetwood2017evolution,fleetwood2022perspective}.
These defects are positively-charged oxygen vacancies ($V_{\rm O}$) of puckered, dimer {(or vacancy cluster with four $E'_\gamma$)}, and forward-oriented configurations~\cite{lenahan1984hole,griscom1984characterization,conley1994observation,yue2017first,wang2024first,griscom1986fundamental,buscarino2005delocalized}.}
{Only $E'_\delta$ and $E'_\gamma$ can exist for irradiation above 200 K~\cite{griscom1984characterization,wang2024first}.}
The dominate $E'_\gamma$ centers can also induce {amphoteric Si dangling bonds ($P_b$ centers) at the SiO$_2$/Si interface}, through a two-stage proton process~\cite{oldham2003total,fleetwood2022perspective}. 
Therefore, revealing the fundamental generation mechanism of $E'_\gamma$ is at the heart of assessing the TID responses of SiO$_2$-based semiconductor devices in harsh radiation environments.

A ``\textit{hole transport-trapping}'' (HTT) mechanism has been proposed 4 decades ago, based on a short-term recovery phenomenon of flat-band voltages in pulse-irradiated metal-oxide-semiconductor (MOS) devices~\cite{oldham2003total,schwank2008radiation,fleetwood2017evolution,fleetwood2022perspective}.
It is suggested that, the $E'_\gamma$ centers are generated because 
the \textit{deep} neutral states can trap holes around them~\cite{oldham2003total,schwank2008radiation}, 
which are transported  via polaron hopping between \textit{shallow} neutral states
under a positive electric field ($E$)~\cite{mclean1976hole,hughes1977time}, 
see Fig. \ref{fig:diagram}(a).
Such an HTT mechanism is supported by previous  non-spin polarized first-principles calculations of single defect levels (SDL)~\cite{lu2002structure,nicklaw2002structure}, which identified 
the shallow and deep localized states as neutral $V_{\rm O}$ of dimer and puckered configurations ($V_{O\delta}$ and $V_{O\gamma}$), respectively.
This $V_{O\gamma}$-based mechanism has been widely applied, not only to 
various scaled MOS devices under Moore's Law~\cite{fleetwood2017evolution,fleetwood2022perspective},
but also to other devices made of alternative dielectrics~\cite{kingon2000alternative} such as HfO$_2$, Ta$_2$O$_5$,
and ion gel~\cite{zhu2023ultra,cramer2016radiation,zhu2020radiation}.
It is so popular that,  
$a$-SiO$_2$ is used as a standard to evaluate the radiation resistance of alternative dielectrics~\cite{felix2002total}.

\begin{figure}[!b]
\centering
\includegraphics[width=0.97\linewidth]{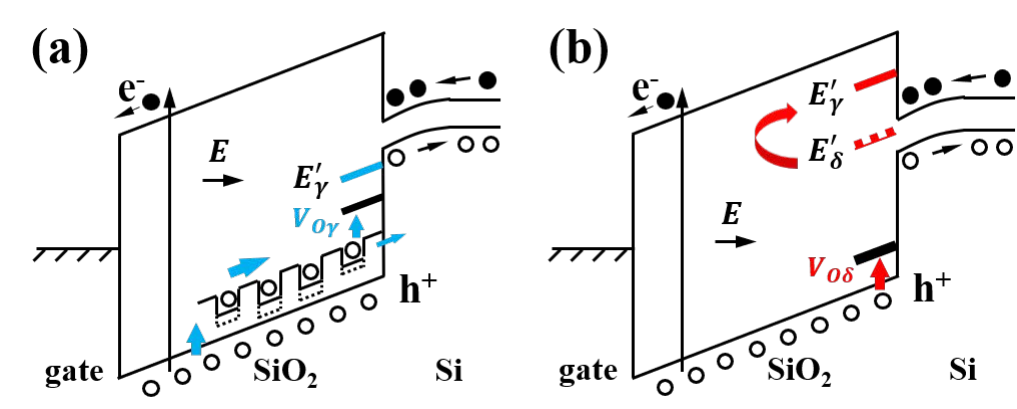}\\ 
\caption{Schematic diagrams of (a) the conventional HTT and (b) our NCCSR mechanisms of $E'_\gamma$ buildup in irradiated $a$-SiO$_2$.
In (a), the partial outflow of holes from the SiO$_2$/Si interface results in the short-term recovery of flat-band voltage.
In (b) with SDLs, $V_{O\gamma}$ are not present, and the stable $E'_\gamma$ are produced from nonradiative hole capture by $V_{O\delta}$ and subsequent structural relaxation of resulting $E'_\delta$.
}
\label{fig:diagram}
\end{figure}

However, this ``standard'' HTT mechanism encounters dilemmas to explain the basic experiments on temperature ($T$) and $E$ dependence of $E'_\gamma$ generation dynamics, when more accurate measurements become available nowadays.
As the trapping of holes by both kinds of $V_{\rm O}$'s are barrierless~\cite{rowsey2011quantitative,hughart2012effects}, 
the defect concentration, $[E'_\gamma]$, should display $T$ and $E$ dependence determined by the much slower hole transport process~\cite{fleetwood2017evolution,fleetwood2022perspective}.
Previous experiments demonstrated that, due to the large transport barrier of $E_b^t\approx 0.4$ eV, 
the time to transport the same number of holes, $t_f\propto e^{E_b^t/k_BT}$ is shortened by nearly 100 times when $T$ is increased from room temperature ($T_R$, 25$^\circ$C) to 150$^\circ$C~\cite{mclean1976hole,hughes1977time}.
However, in recent experiments only a 3-fold shortening is observed in the time to generate the same $[E'_\gamma]$ as $T$ is elevated in the range~\cite{dong2020evolution,witczak1997hardness}. 
Moreover, previous experiments also show that, the $t_f$ decreases sharply as $E$ increases because the hole transport is accelerated~\cite{mclean1976hole,hughes1977time}.
However, recent experiments show that $[E'_\gamma]$  
displays a first increase and then decrease behavior as $E$ increases~\cite{shaneyfelt1990field}.
These much weaker $T$ and opposite $E$ dependence of $E'_\gamma$ buildup relative to the hole transport suggest that, the conventional HTT mechanism proposed from the SDLs is hardly the origin of the $E'_\gamma$ in ionizing-irradiated $a$-SiO$_2$. 
Note that, as a WBG glass, significant metastability and dispersion can exist in the strongly localized defect electronic states of $a$-SiO$_2$~\cite{wilhelmer2022ab,qiu2023dual,wang2024first,guo2024si},
and the SDL method may fail to identify the defect generation mechanism.

In this Letter, we propose a new  ``nonradiative carrier capture-structural relaxation'' (NCCSR) mechanism based on an FECCC [formation energy ($\Delta H_f$) and capture cross-section ($\sigma$)] method  
to justify the defect thermal stabilities and carrier capture paths of $V_{\rm O}$'s in $a$-SiO$_2$. 
{Our calculations using {HSE06 hybrid functional~\cite{heyd2003hybrid}}  
yield novel defect parameters  
aligning with the EPR and carrier injection experiments~\cite{lenahan1984hole,buscarino2009polyamorphic,conley1994observation_JAP,conley1994observation},
but significantly deviate from 
previous calculations using PBE and PBE0 functionals~\cite{yue2017first,wang2024first,wilhelmer2022ab}.}
Our analysis indicates that, the long-assumed $V_{O\gamma}$ precursors with large $\Delta H_f$  
cannot survive in high $T$-grown $a$-SiO$_2$; instead, the stable $V_{O\delta}$ can capture irradiation-induced holes in 
the valence band (VB) of $a$-SiO$_2$ even at $T_R$, 
and {about 4/5 of the resulting $E'_\delta$} are metastable thus 
can deform to the stable $E'_\gamma$, see Fig.~\ref{fig:diagram}(b).
We demonstrate that, such an NCCSR mechanism can consistently explain the puzzling $T$ and $E$ dependences. 
{By using the rigorous Kohlrausch-Williams Watts (KWW) decay function~\cite{kohlrausch1854theorie,williams1970non} to overcome the difficulty of determining the distribution in the traditional integral~\cite{grasser2008dispersive},}
we also successfully derive {a fractional power law (FPL) dynamic model} to uniformly predict the sublinear experimental data over a wide dose and $T$ range.

{\textit{The FECCC method.}--}
Relative to the distribution of SDLs, charge transition levels ($E_t$) and $\sigma$'s are more relevant to justify the charge capture paths 
of the defects and precursors in $a$-SiO$_2$.
Meanwhile, $\Delta H_f$ and structural relaxation barriers ($E_b^r$) should be used to analyze  
the thermal stabilities of the defect and precursor electronic states.
{Moreover, each $V_{\rm O}$ site is different in  $a$-SiO$_2$ and all these quantities should be studied in a statistical manner.}
Considering all these factors, here, we adopt an FECCC method to use the distribution of $\Delta H_f$, $E_b^r$, $E_t$, and $\sigma$
to identify the generation paths of $E'_\gamma$ in ionizing-irradiated $a$-SiO$_2$.

Our first-principles calculations of electronic structure and total energy are carried out using \emph{spin-polarized} density-functional theory (DFT), as implemented in the PWmat package~\cite{wang2019some} 
with the NCPP-SG15-PBE pseudopotentials~\cite{jia2013fast,jia2013analysis,schmidt2012electrostatic}. 
To improve the accuracy, the Heyd-Scuseria-Ernzerhof (HSE06) hybrid functional~\cite{heyd2003hybrid} with a mixing parameter of 50\% is employed {to achieve the band gap of 9 eV}. 
To study the statistical properties of $V_{\rm O}$'s, we followed Shluger and coauthors’ pioneering work, in which many 216-atom amorphous supercells are created by melt-and-quench procedure of crystalline SiO$_2$ {and compared with experiments to ensure good representatives of $a$-SiO$_2$~\cite{el2014nature,zhu2022effect}.}
In this work, we take the first 10 amorphous supercells and randomly selected 20 samples of $V_{\rm O}$ sites to include the effects of the amorphous nature.
All atoms within the supercell are relaxed until the forces on each atom fall below 0.01 eV/Å and the plane-wave energy cutoff for the basis-functions is set to be 60 Ry.

{The $\Delta H_f$, $E_b^r$, $E_t$  
are calculated by conventional methods, see the Supplementary Material (SM)~\cite{SM} for details.}
To obtain the $\sigma$ of $V_{\rm O}$'s, the electron-phonon coupling constants ($C_{ij}^k$) between electronic states $i$ and $j$, as well as phonon mode $k$ are calculated using the static coupling approximation~\cite{passler1982nonradiative} 
and are obtained all at once by a novel single self-consistent field approach proposed by Wang et al.~\cite{shi2012ab,shi2015comparative}.
{For $T>$300 K, $\sigma$ can be written in term of $C_{ij}^k$ and an effective thermal barrier ($E_b^{ij}$)~\cite{alkauskas2014first,huang1981lattice},} 
\begin{equation}
\sigma= W_{ij}V\sqrt{\frac{m^*}{3k_BT}} = 
V\sqrt{\frac{\pi m^*}{3\lambda_{ij}}}\left ( \sum_k |C_{ij}^k|^2/\omega_k^{2} \right) e^{-\frac{E_b^{ij}}{k_BT}}.
\label{eq:cross-section}
\end{equation}
Here $W_{ij}$ is non-radiative decay probability, 
$V$ is the supercell volume, and $m^*=$0.58 $ m_e$ (0.3 $m_e$) is the effective mass of hole (electron) in $a$-SiO$_2$ ($m_e$ is the mass of free electron)~\cite{kumar2011determination}.
$\lambda_{ij}$ is the structural relaxation energy between initial and final configurations after the transition from $i$ to $j$; 
$\omega_k$ is the frequency of the $k$-th harmonic phonon mode.
$E_b^{ij}=(E_i-E_j-\lambda_{ij})^2/4 \lambda_{ij}$ is the thermal barrier for crossing to different charge states of the same configuration.

\begin{figure}[!t] 
\centering
\includegraphics[width=0.95\linewidth]{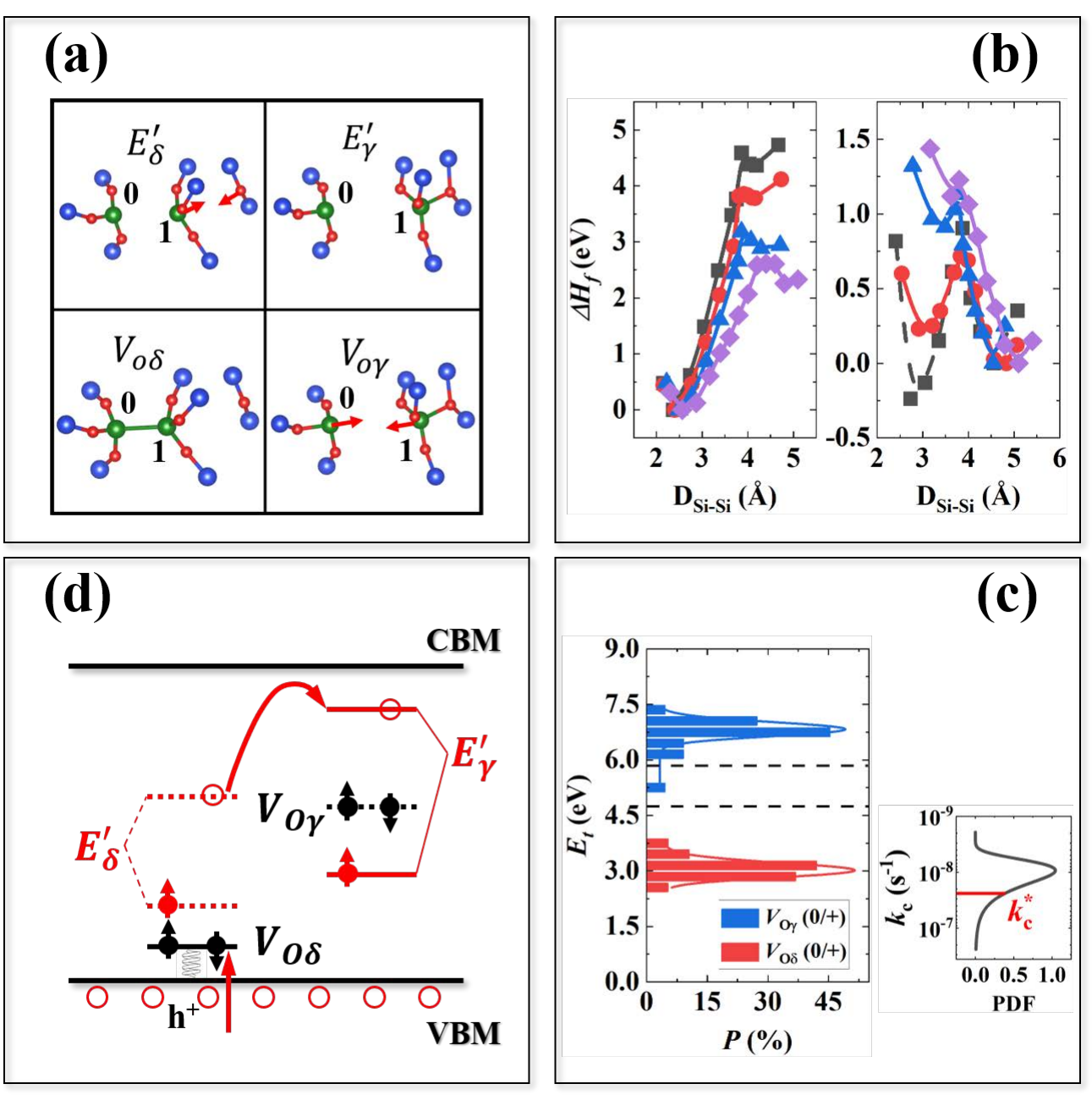} 
\caption{Basis for our NCCSR mechanism.
(a) Structures of the two configurations of $V_{\rm O}$'s at neutral (0) and positive (+) charged states.
Bigger (smaller) balls are for Si (O) atoms.
(b) Defect formation energies of $V_{\rm O}^0$ (left column) and $V_{\rm O}^+$ (right column) as a function of the distance between Si0 and Si1 denoted in (a).
(c) Left: distributions of $(0/+)$ transition levels of the dimer (red) and puckered (blue) configurations.
Right: asymmetric distribution of $k_c$ calculated from 
the KWW decay function.
(d) SDLs of the four defect states in (a). 
Our NCCSR mechanism is indicated by the two red arrows. 
}
\label{fig:basis}
\end{figure}

{\textit{Calculation results and the NCCSR mechanism.}--}
The obtained structures of a typical sample are presented in Fig.~\ref{fig:basis}(a).
It is clear that, {consistent with previous calculations~\cite{boero1997structure,blochl2000first,wilhelmer2022ab,wang2024first},} $V_{\rm O}$'s in $a$-SiO$_2$ can take two distinct local configurations for each charge state, namely dimer $V_{O\delta}$ and puckered $V_{O\gamma}$ for neutral state, and $E’_\delta$ and $E’_\gamma$ for positively charged state.
{The $V_{\rm O}$ is amphoteric~\cite{wilhelmer2022ab}, but the ($0/-$) transition levels are too deep to have significant impacts on the TID effects.}

According to the FECCC method, we first investigate the thermal stabilities of $V_{\rm O}$'s in $a$-SiO$_2$ based on $\Delta H_f$ and $E_b^r$.
The difference in $\Delta H_f$ between the dimer and puckered configures, $\Delta E$, are shown in Fig. S1 {in the SM~\cite{SM}.}
{In Fig. S1 (a), all $V_{O\gamma}$ are found to have higher energies than the corresponding $V_{O\delta}$, with a difference $\Delta E_0$ of $-4.4\sim-1.8$ eV.
In Fig. S1 (b), for about 4/5 (1/5) of the samples, $E'_\gamma$ have lower (higher) $\Delta H_f$ than the $E'_\delta$ configuration, with a difference $\Delta E_+$ of $-0.3$$\sim$1.6 eV.}  
{Fig. S2 in the SM~\cite{SM} shows that, the $\Delta E_+$ have a near-linear correlation with  
$-\Delta E_0$.
Therefore, 4 samples (denoted by the red dots) were selected} to calculate $\Delta H_f$ as a function of the distance between Si0 and Si1 as denoted in Fig. 2(a).
The results are present in Fig.~\ref{fig:basis}(b).
{It is found that, the greater the $-\Delta E_0$ ($\Delta E_+$), the smaller the $E_b^r$ of $V_{O\gamma}\rightarrow V_{O\delta}$ ($E'_\delta\rightarrow E’_\gamma$), 
which ranges in 0.09$\sim$0.40 eV (0.013$\sim$0.54 eV).} 
Assuming an attempt frequency of 10$^{13}$s$^{-1}$~\cite{wimmer2016role}, 
we estimate that {the time  
to transform from metastable $V_{O\gamma}$ to stable $V_{O\delta}$ ranges in $3.2\times10^{-3}\sim 4.8\times10^2$ ns, and to transform from metastable $E’_\delta$ to stable $E’_\gamma$ ranges in $1.1\times10^{-4}\sim1.0\times10^{5}$ ns.}
The latter 
has been observed in previous experiments~\cite{warren1994microscopic}.

{These statistics justify the stability of defect and precursor states.
First, 
almost all ground states of the neutral precursors are $V_{O\delta}$, instead of the long-assumed $V_{O\gamma}$~\cite{oldham2003total,schwank2008radiation}.}
This is because, the $a$-SiO$_2$ dielectrics in modern semiconductor devices are prepared by thermal oxidation process  
at typical high temperature of 800$^\circ$C$\sim$1100$^\circ$C.
Thus, the relatively small relaxation barriers of $V_{O\gamma}\rightarrow V_{O\delta}$ can all be overcome, and the metastable
$V_{O\gamma}$ \textit{can hardly exist} in $a$-SiO$_2$.
As seen in Fig. 2(a), the structural relaxation occurs because the metastable $V_{O\gamma}$ have dipole structures~\cite{lu2002structure}; Si0 with an electron and Si1 with a hole can attract each other and bond to form the stable $V_{O\delta}$.
{Second, 
about 4/5 (1/5) of the ground states of positively-charged defects are $E’_\gamma$ ($E'_\delta$).}
{This ratio is essentially different from $E'_\delta$-dominated ground state in previous studies using PBE0 functional~\cite{wilhelmer2022ab}, but is} consistent with EPR experiments~\cite{lenahan1984hole,buscarino2009polyamorphic}.
As seen in Fig. 2(a),
Si1 in metastable $E'_\delta$ feels Coulomb attraction from neighboring O ions, and can relax through the base plane of three O atoms and bonds with another network O atom, forming the stable $E'_\gamma$.
These results on ground states imply that, $E'_\gamma$ are unlikely to be generated from the HTT mechanism considering $V_{O\gamma}$ as precursors.

According to the FECCC method, 
the charge capture paths of $V_{\rm O}$'s are further investigated based on $E_t$ and $\sigma$.
In Fig.~\ref{fig:basis} (c), the distributions of $E_t$ of the dimer and puckered configurations 
are found at 2.4$\sim$3.9 eV and 6.0$\sim$7.5 eV above the valance band maximum (VBM), respectively.
{These $E_t$'s are much higher than previous calculations using spin polarized PBE or non-spin polarized PBE0 functional~\cite{wilhelmer2022ab,wang2024first,yue2017first}.}
{Fig. S3 in the SM~\cite{SM} shows that, the discrepancy originated from underestimating the bandgap and not considering spin polarization for atomic structure relaxation in the previous calculations.}
{Because the computational cost to include all phonon modes is large, the $\sigma$ are calculated for the four selected samples.}
{The stable $V_{O\delta}$'s are found to have large $\sigma_{\delta,h}=5.4\times10^{-12}\sim 1.5\times10^{-10}{\rm cm}^2$ at $T_R$,
{which are much larger than previous calculations~\cite{yue2021first},
but close to carrier injection experiments~\cite{conley1994observation,conley1994observation_JAP}.}
The underlying parameters in Eq. (1) are listed in Tab. I in the SM~\cite{SM}.}
$\sigma_{\delta,h}$ are large not only because $V_{O\delta}$ have relatively shallow
$E_t$ (2.5$\sim$3.5 eV above the VBM) thus strong $C_{ij}^k$ between $V_{O\delta}$ and $E'_\delta$ electronic states,
but also because they have large {$\lambda_{ij}(=$1.7$\sim$2.5 eV}) comparable to the $E_t$'s, which leads to small hole capture barrier ($E_b^{hc}$) of 0.095$\sim$0.11 eV.
{These nonzero barriers are essentially different from previous barrierless trapping~\cite{rowsey2011quantitative,hughart2012effects}.}
The metastable $E’_\delta$ states are found to have negligible $\sigma_{\delta,e}$ smaller than $10^{-20}{\rm cm}^2$, 
because the $E_t$'s are too deep [{5.5$\sim$6.5 eV below the conduction band minimum (CBM)}] to capture electrons from the CB of $a$-SiO$_2$.
Meanwhile, the stable $E'_\gamma$ are found to have  {$\sigma_{\gamma,e}=2.4\times10^{-15}\sim 4.5\times10^{-14}{\rm cm}^2$} at $T_R$, {which are consistent with carrier injection experiments~\cite{conley1994observation,conley1994observation_JAP}}.
The underlying parameters in Tab. II in the SM~\cite{SM} show that, $\sigma_{\gamma,e}$ are hundreds of times smaller than $\sigma_{\delta,h}$, mainly because $\lambda_{ij}$ become much smaller (1.0$\sim$1.1 eV) and the $E_b^{ec}$ becomes much larger ({0.22$\sim$0.27 eV}).
The metastable $V_{O\gamma}$ states have negligible $\sigma_{\gamma,h}$ smaller than $10^{-20}{\rm cm}^2$, because their $E_t$'s ({6.8$\sim$7.0 eV}) are far away from the CBM.
{Due to $a$-SiO$_2$'s pivotal role in  MOS devices, optical fibers, and memories~\cite{kingon2000alternative}, these novel defect parameters  
can support broad $V_{\rm O}$-related reliability and function research.}

The NCCSR mechanism emerges from these analysis on defect ground states and charge capture paths.
It is illustrated in Fig.~\ref{fig:basis}(d), in which SDLs of the sample denoted by the first red dot in Fig. S2 in the SM~\cite{SM} are plotted.
Due to the large $\sigma_{\delta,h}$ and negligible $\sigma_{\delta,e}$, the stable $V_{O\delta}$ can easily capture irradiation-induced holes in the SiO$_2$ VB even at $T_R$, and the resulting $E’_\delta$ centers will not turn back by a recombination process.
With larger formation energies and small relaxation barriers, {most of them will deform to the stable $E'_\gamma$ centers,} 
\begin{equation}
V_{O\delta} + h^+ {\rightarrow} E'_\delta \rightarrow E'_\gamma.
\label{eq:mechanism}
\end{equation}
{The few others will stay as the ground states of $V_{\rm O}^+$, i.e., $V_{O\delta} + h^+ {\rightarrow} E'_\delta$.}
As the CB electrons are much fewer than the VB holes~\cite{oldham2003total,schwank2008radiation} and $\sigma_{\gamma,e}\ll\sigma_{\delta,h}$, the generated $E'_\gamma$ are hard to be eliminated by the induced electrons.
This new NCCSR mechanism is also demonstrated in Fig.~\ref{fig:diagram}(b) to compare with the conventional HTT mechanism in Fig.~\ref{fig:diagram}(a).

{\textit{Dynamic model of $E'$ generation.}--}
To link the NCCSR mechanism to TID irradiation experiments, we derive an FPL model of $E'$ generation dynamics {based on 
the {observed amorphous nature} of $a$-SiO$_2$.
Ionizing irradiation drives a collective exponential relaxations of dispersive $V_{O\delta}$'s in $a$-SiO$_2$~\cite{johnston2006stretched}.
Therefore, the annihilation dynamics of $V_{O\delta}$ 
can be obtained by integral of $[V_{O\delta}](t)=\int_0^\infty P(k_c) e^{-k_ct}dk_c$~\cite{grasser2008dispersive},} where $k_c=v_{h}\sigma_{\delta,h}p_0$ is the rate constant of hole capture~\cite{huang1981lattice}, $v_{h}\propto\sqrt{T}$ and $p_0$ are the thermal velocity and density~\cite{johnston2014field} of holes in the SiO$_2$ VB, respectively.
$P$ is the probability distribution function (PDF) and $\int_0^\infty P(k_c) dk_c = D_0$ is the initial concentration of $V_{O\delta}$.
Due to the annihilation-formation correspondence, the concentration of $E'$ can be obtained by
$[E'](t)=D_0-[V_{O\delta}](t)$.

However, obtaining the accurate PDF of $k_c$ is not practical, 
{due to the deviation of calculation results from the actual situation and/or the mismatch between the selected PDF and the calculation results.}
We notice that, 
{numerous studies have demonstrated that~\cite{berthier2011theoretical,phillips1996stretched,
kakalios1987stretched,kriza1986stretched,mihaly1991dielectric,johnston2005dynamics},} collective relaxation in amorphous materials 
follow a rigorous KWW decay fuction~\cite{kohlrausch1854theorie,williams1970non}.
{In this work, we pioneer its application to hole capture-induced relaxation of $V_{O\delta}$ in ionizing-irradiated $a$-SiO$_2$.}
The concentration reads  
$[V_{O\delta}](t)=D_0 e^{-(k_c^*t)^\beta}$, instead of a conventional Debye exponential form ($\beta=1$). 
The PDF of $k_c$ can be obtained through an inverse Laplace transform of the stretched exponential~\cite{lindsey1980detailed,berberan2005mathematical}, in which, $k_c^*$ and $\beta$ are found as the dividing point for about half of the distribution and a measure of the inherently small rate cutoff of the PDF, respectively~\cite{johnston2006stretched}.
The generation dynamics of $E'$ is given by
$[E'](t)= D_0 [1-e^{-(k_c^*t)^\beta}]$.
Considering that very few $V_{O\delta}$ are consumed~\cite{blochl2000first},
we perform a first-order Taylor expansion of the result about $t=0$, and obtain an FPL model of the $E'$ generation dynamics in irradiated $a$-SiO$_2$,
\begin{equation}
[E'](t)= D_0 (k_c^*t)^\beta = \kappa_c D^\beta.
\label{eq:model}
\end{equation}
Here $\kappa_c=D_0 (k_c^*/R)^\beta$, $R$ is the dose rate, and $D$ the total dose.

This analytic formula  is
{much concise than the conventional numerical models consisting of a dozen coupled differential equations~\cite{rowsey2011quantitative,hughart2012effects,xu2017multi}}, 
and {naturally explains the universal FPL behavior observed in experiments~\cite{griscom1993radiation,mashkov1996fundamental}}.
It is compared with previous experimental data~\cite{dong2020evolution} to demonstrate its universality.
Fig.~\ref{fig:demonstrate}(a) shows that,
the formula can uniformly describe the nonlinear data over a wide $T$ and $D$ range.
The parameters in Eq. (\ref{eq:model}) are extracted from the unified fitting.
$\beta$ is found to be close to 2/3 for all $T$~\cite{winokur1980interface},
which is much larger than the dispersion factor of $\sim$0.22 for hole transport in $a$-SiO$_2$~\cite{mclean1976hole,hughes1977time}, further reflecting the irrelevance of the HTT model. 
In Fig.~\ref{fig:demonstrate}(b), $k_c^*$ is obtained by setting $D_0$ as $1.0\times10^{14}$cm$^{-2}$~\cite{blochl2000first}, and is found to display an Arrhenius-like $T$ dependence.
(It becomes smaller for larger $D_0$, but the $T$ dependence is almost unaffected.)
Considering the definitions of $k_c^*$ and $\sigma_{\delta,h}$, $V_{O\delta}$'s hole capture barrier can be calculated by $e^{-E_b^{hc}/k_BT}\propto k_c^*(T)/\sqrt{T}$.
The obtained $E_b^{hc}=$0.11 eV agrees with our first-principles calculation results (0.095$\sim$0.11 eV).
{The PDF of $k_c$ solved with $k_c^*$ and $\beta$ [right column in Fig. 2(c)] displays an asymmetric distribution similar to that of the transition levels [left column in Fig. 2(c)].}
{Therefore, our physics-driven KWW modeling approach overcomes the difficulty of determining the distribution in the traditional integral, enables unified and predictive defect dynamics modeling in amorphous dielectrics and semiconductors.}

\begin{figure}[!t]
\centering
\includegraphics[width=0.97\linewidth]{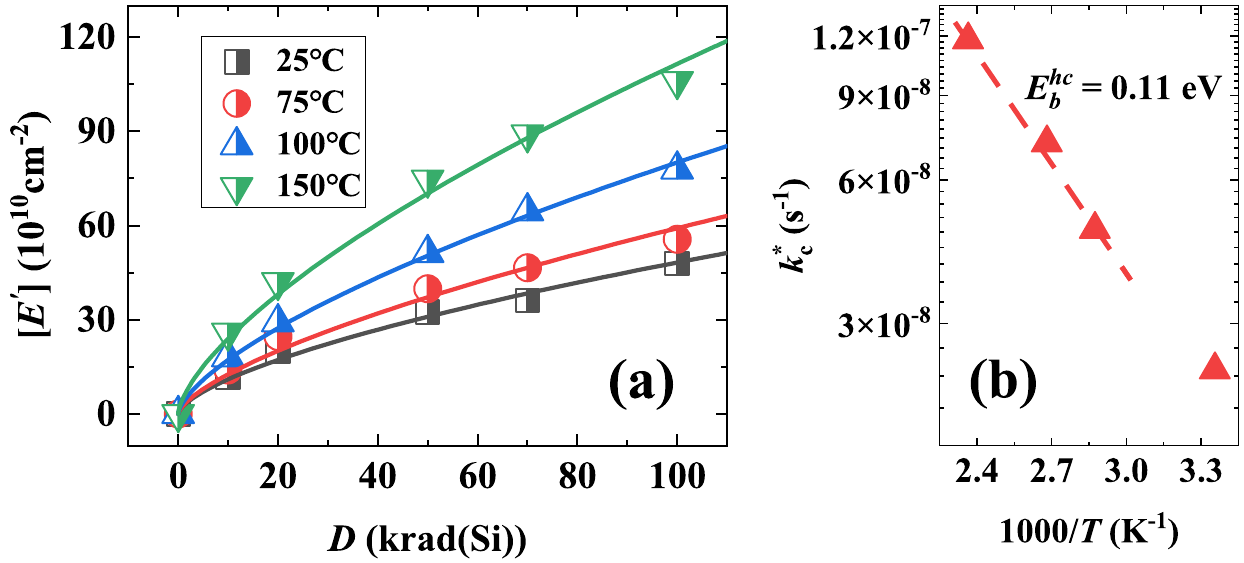}\\ 
\caption{(a) Experimental data (dots) and fitting results (curves) of 
$[E']$ as a function of $D$ at different $T$. 
(b) Extracted $k_c^*$ as a function of $1/T$.
The data in (a) are the sum of measured 
$[P_b]$ and $[E']$ in MOS structures irradiated by $^{60}$Co $\gamma$-ray at 10 rad(Si)/s and zero bias~\cite{dong2020evolution}. 
This is reasonable due to the one-to-one transformation~\cite{chang1986amphoteric,stahlbush1993post} and negligible annealing effects~\cite{dong2020evolution} of the two kinds of defects.}
\label{fig:demonstrate}
\end{figure}

{\textit{Origins of the $T$ and $E$ dependences.}--}
We now demonstrate that the basic temperature~\cite{dong2020evolution,witczak1997hardness} 
and electric field~\cite{shaneyfelt1990field} dependences of $E'$ buildup dynamics in $a$-SiO$_2$, which cannot be understood by the conventional HTT mechanism, can be readily explained by our proposed NCCSR mechanism.
From Eq. (3), $[E'](t)\propto (v_{h}\sigma_{\delta,h}p_0t)^\beta$.
As $p_0$ weakly depends on $T$~\cite{taylor1984geminate}, the $T$ dependence of $[E']$ at a fixed $t$ is determined by 
$v_{h}(T)\sigma_{\delta,h} (T) \propto \sqrt{T}{\rm Exp}(-E_b^{hc}/k_BT)$.
From the calculated and fitted small $E_b^{hc}\approx0.11$ eV,
it is readily derived that the time to produce defects of the same concentration decreases only 2.3 times as $T$ increases from $T_R$ to 150$^\circ$C.
This evaluation is very close to the experimental observations of defect dynamics~\cite{dong2020evolution,witczak1997hardness}, 
but is much weaker than the near 100-fold shortening of $t_f$ derived in the hole transport dynamics~\cite{mclean1976hole,hughes1977time}. 
On the other hand, as $v_h$ is independent of $E$, $[E']$ at a fixed time displays $E$ dependence as a product of $p_0(E)$ and $\sigma_{\delta,h}(E)$.
$p_0$ has a positive dependence on $E$~\cite{johnston2014field}, because the spatial separation of irradiation-induced electron-hole pairs becomes more significant under a stronger electric field. 
The $\sigma_{\delta,h}$ of $V_{O\delta}$ has a negative dependence on $E$, because an electric field induces a tilt of the hole-induced polarization potential of the neutral $V_{O\delta}$, whose volume
below a critical value is proportional to $\sigma_{\delta,h}$~\cite{dussel1970field,lax1960cascade}.
Therefore, $[E']$ as a product of $p_0(E)$ and $\sigma_{\delta,h}(E)$ display a first increase and then decrease behavior at relatively small and large $E$, respectively.
This basic trend is consistent with the observation in defect generation dynamics~\cite{shaneyfelt1990field}, but is essentially different from the monotonic $E$ dependence observed in hole transport dynamics~\cite{mclean1976hole,hughes1977time}.

{\textit{{Role of the metastability and dispersion}.}--}
The SDL distribution are shown in Fig. S4 in the SM~\cite{SM}, from which the relatively shallow $E'_\delta$ and deep $E'_\gamma$ seem to support the HTT mechanism,
which however, cannot explain the basic $T$ and $E$ dependence 
~\cite{dong2020evolution,witczak1997hardness,shaneyfelt1990field}. 
Our FECCC analysis show clearly that, this is because  
the SDLs cannot characterize the \textit{metastability} 
of strongly-localized defect electronic states in $a$-SiO$_2$ and other WBG materials~\cite{park1999first,park2002origin,li2005stability}. 
{The \textit{dispersion} of them is also indispensable in
our KWW modeling of non-equilibrium defect generation dynamics.
It was not discussed in the previous study on nonradiative carrier recombination in WBG materials~\cite{qiu2023dual}.}
{Both the electronic metastability and dispersion play critical roles in the understanding of the bias temperature instability (BTI) effects~\cite{wimmer2016role,wilhelmer2022ab}.  
However, our NCCSR model of TID (generation of $E'$ by capturing holes in SiO$_2$ VB even at $T_R$ and zero bias) is essentially different from the four-state model of BTI~\cite{wimmer2016role,wilhelmer2022ab}, which considers
the generation of hydroxyl-$E'$ by trapping carriers from the Si substrate at high temperature and large bias.}
As electronic metastability and dispersion exist extensively in alternative dielectrics (amorphous HfO$_2$, Al$_2$O$_3$,  BN, etc~\cite{choi2009charge,momida2020bistability,hong2020ultralow}), our FECCC-KWW approach can be applied to study the radiation damage physics of these materials.

{\textit{Conclusion.}--}
In conclusion, to solve the puzzle of the experimentally observed $T$ and $E$ dependence of $E'_\gamma$ formation in ionizing-irradiated $a$-SiO$_2$, 
we have investigated the defect generation mechanism by using the FECCC analysis and  KWW modeling methods.
{Based on novel defect parameters  
aligning with EPR and carrier injection experiments,} 
we have proposed the NCCSR formation mechanism and FPL dynamic model for $E'_\gamma$ formation in ionizing-irradiated $a$-SiO$_2$,
which can consistently explain and uniformly describe  
experimental data over wide dose, temperature, and electric field ranges.
Our groundbreaking mechanism and model should {serve as a cornerstone  
of TID effects} of SiO$_2$-based semiconductor devices.
The FECCC-KWW methodology is expected to be general for alternative dielectrics 
with intrinsic amorphous nature and electronic metastability.

This work was supported by the Xinjiang Tianchi Talents Program,
the National Natural Science Foundation of China (Grant No. {12447158}), and 
the National Key Research and Development Program of China (Grant No. 2024YFA1409800).

\end{sloppypar}


%

\appendix

\renewcommand{\thefigure}{S\arabic{figure}}
\setcounter{figure}{0}

\newpage

\title{Supplementary Material for ``Mechanism of $E'_\gamma$ Defect Generation in Ionizing-irradiated $a$-SiO$_2$: The Nonradiative Carrier Capture-Structural Relaxation Model''} 

\author{Yu Song} 
\affiliation{
Xinjiang Key Laboratory of Extreme Environment Electronics, Xinjiang Technical Institute of Physics and Chemistry, Chinese Academy of Sciences, Urumqi 830011, China}
\altaffiliation{Previous at Neijiang Normal University, Neijiang 641112, China}

\author{Chen Qiu}
\affiliation{Eastern Institute of Technology, Ningbo 315200, China}

\author{Su-Huai Wei$^*$}
\email{suhuaiwei@eitech.edu.cn}
\affiliation{Eastern Institute of Technology, Ningbo 315200, China}

\date{May 13, 2025}

\maketitle

\newpage
{\bf \textit{Calculation methods of ${\Delta}{H}_{f}$, ${E}_{b}^{r}$, and ${E}_{t}$.}--}
The formation energy ($\Delta H_f$) of a defect $\alpha$ at charge state $q$ is calculated by $\Delta H_f(\alpha, q) = \Delta E(\alpha, q) + \sum_i n_i \mu_i + qE_F$, where $\Delta E(\alpha, q)=E(\alpha, q)-E(\rm{host}) + \sum_{i} n_{i}E_{i} + qE_{\rm{VBM}}^{\rm{host}}$  [S1]. Here $E\left(\alpha,q\right)$ is the total energy of a supercell containing the relaxed defect $\alpha$ at charge state $q$, $E\left(host\right)$ is the total energy of the same supercell in the absence of the defect, and $E_{VBM}^{host}$ is the energy of host VBM. $n_i$ denotes the number of atoms removed from the host supercell to form supercell containing the defect; $E_i$ represent the energy of the elemental stable solid/gas and $\mu_i$ is the chemical potential of each of the components $i$. $q$ describes the number of electrons transferred from the supercell to the reservoirs and $E_f$ is the Fermi level. The structural relaxation barrier ($E_b^r$) between defects of the same charge state but different configurations is usually calculated by the Nudged Elastic Band (NEB) method [S2]. However, we have found that this method is almost non-convergent for our amorphous structures. Therefore, we adopt the discrete path sampling (DPS) method [S3], in which all degrees of freedom except for the Si-Si distances are allowed to be relaxed, and the energy of a series of intermediate configurations are calculated to estimate the relaxation barrier. The charge transition level ($E_t$) is defined as the $E_F$ at which the formation energy for defect with charge state $q$ and $q'$ are equal, $E_t(q/q') = \left[ \Delta E(\alpha, q) - \Delta E(\alpha, q') \right] / (q' - q)$.

{\bf References:}\\
$[\rm{S1}]$ S.-H. Wei, Comp. Mater. Sci. {\bf 30}, 337 (2004).\\
$[\rm{S2}]$ H. Jonsson, G. Mills, and K. W. Jacobsen, in Classical and quantum dynamics in condensed phase simulations (World Scientific, 1998) pp. 385–404.\\
$[\rm{S3}]$ D. J. Wales, Mol. Phys. {\bf 100}, 3285 (2002).

\clearpage
\begin{figure}
    \centering
    \includegraphics[width=0.75\linewidth]{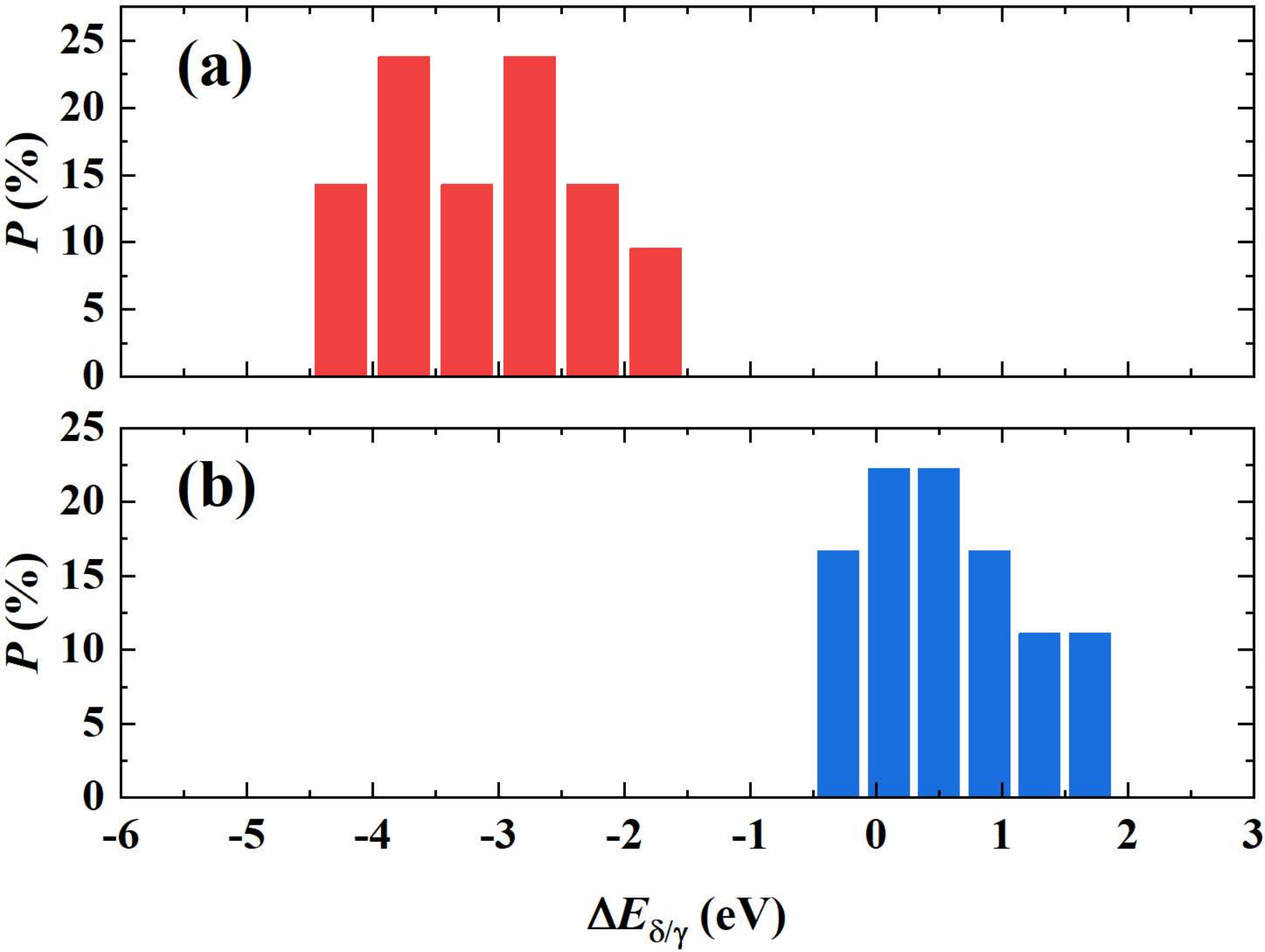}
    \caption{Distribution probability of the difference in formation energies between the dimer and puckered configures ($\Delta E_{\delta/\gamma}$) for (a) neutral and (b) positively-charged V$_O$’s.}
    \label{fig:enter-label}
\end{figure}

\clearpage
\begin{figure}
    \centering
    \includegraphics[width=0.65\linewidth]{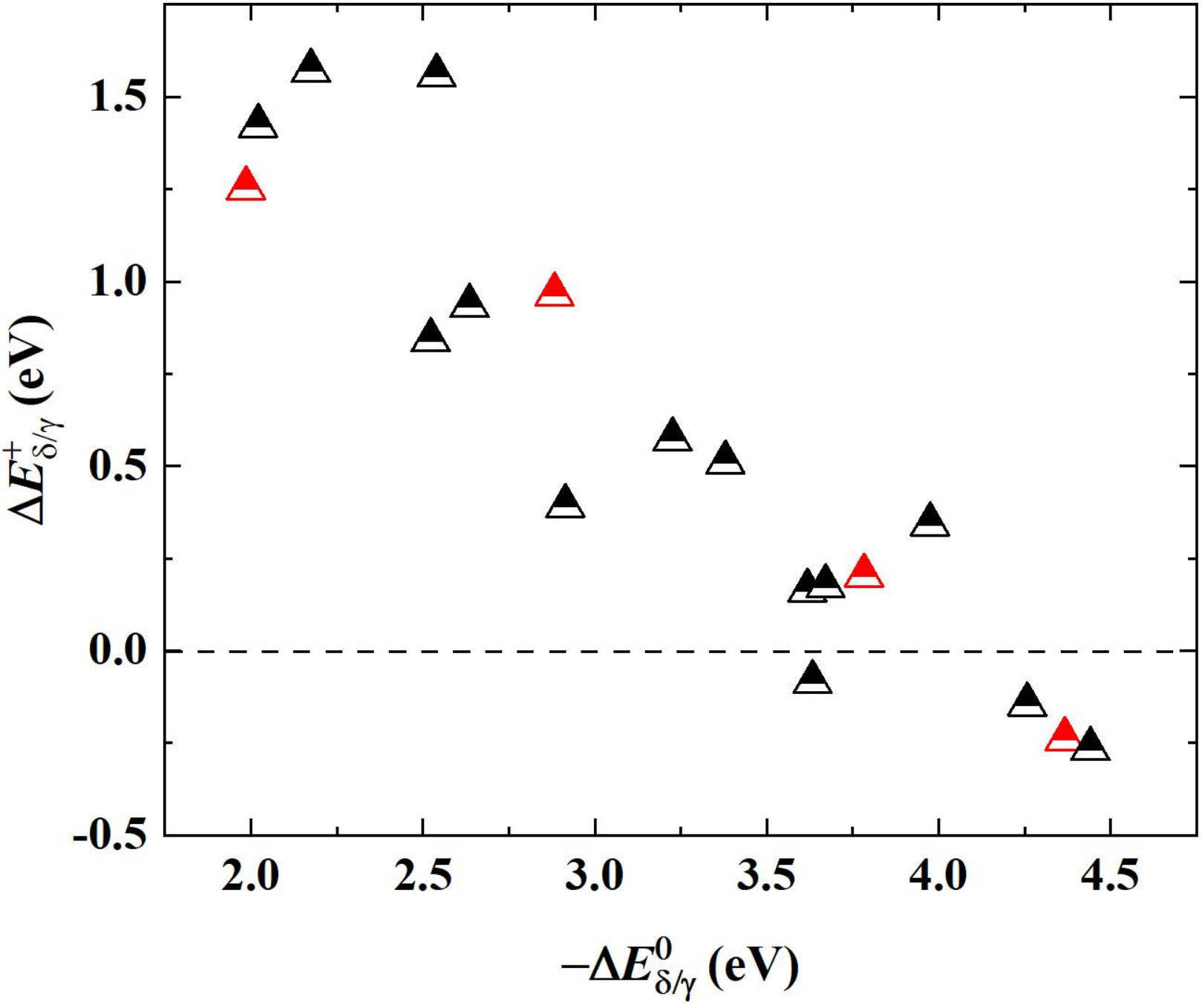}
    \caption{Near-linear correlation between $\Delta E_{\delta/\gamma}^+$ and $-\Delta E_{\delta/\gamma}^0$ in Fig. S1.}
    \label{fig:enter-label}
\end{figure}

\clearpage
\begin{figure}
    \centering
    \includegraphics[width=0.95\linewidth]{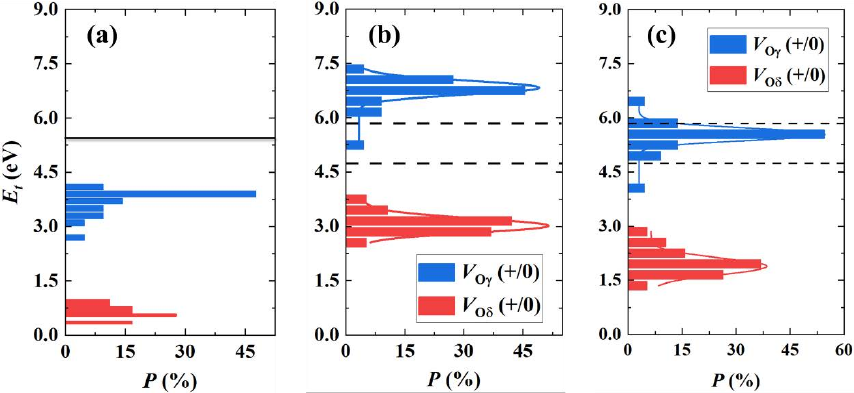}
    \caption{Distribution probabilities ($P$) of charge transition levels calculated by using (a) spin-polarized PBE functional, (b) spin-polarized HSE06 functional, and (c) HSE06 without spin polarization. In (a) the band gap of SiO$_2$ is only 5.44 eV (about 2/3 of experimental value), and the $E_t$’s of both the dimer and puckered configurations are much lower than our calculations using spin polarized HSE06 in (b). Using spin polarization for atomic structure relaxation result in a much different configuration from not considering spin polarization, leading to differences in $E_t$’s in (b) and (c).}
    \label{fig:enter-label}
\end{figure}

\clearpage
\begin{figure}
    \centering
    \includegraphics[width=0.75\linewidth]{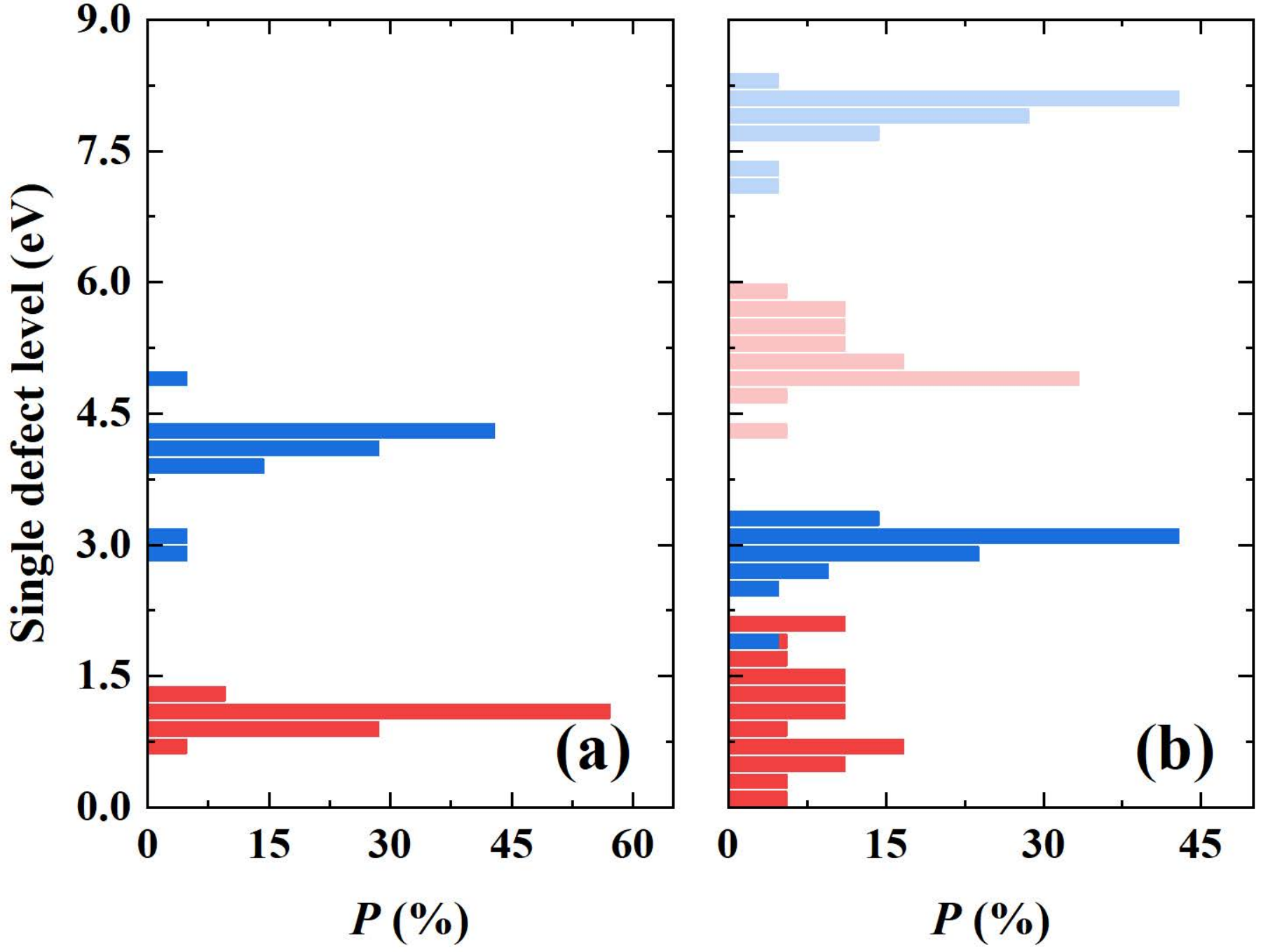}
    \caption{Single defect level distributions of (a) neutral and (b) positively-charged V$_O$’s. In (a), the occupied levels of $V_{O\delta}$ and $V_{O\gamma}$ are located at 0.6$\sim$1.4 eV (red) and 2.8$\sim$5.0 eV (blue) above the VBM of $a$-SiO$_2$, respectively. In (b), $E'_\delta$ and $E'_\gamma$ have empty single electron energy states of 3.0$\sim$4.8 eV (light red) and 0.6$\sim$2.2 eV (light blue) below the CBM of $a$-SiO$_2$, respectively. Their occupied single electron energy states are located at 0$\sim$2.4 eV (red) and 1.8$\sim$3.4 eV (blue) above the VBM of $a$-SiO$_2$, respectively.}
    \label{fig:enter-label}
\end{figure}

\clearpage
\begin{table}[t]
    \centering
    \caption{{Calculated hole capture cross-section of ${V}_{{O\delta}}$ and underlying parameters.}}
    \label{tab:my_label}
    \begin{tabular}{ccccc}
    \toprule
        {\bf Sample No.} & {\bf 1} & {\bf 2} & {\bf 3} & {\bf 4}\\
    \midrule
        ${\sigma}_{{\delta},{h}}$ (${{10}}^{-{11}}{\rm cm}^2$) & 1.60 & 0.54 & 15.3 & 1.63\\
        ${E}_{{ij}}$ (eV) & 3.50 & 3.09 & 2.79 & 2.54\\
        ${\lambda}_{{ij}}$ (eV) & 2.52 & 2.11 & 1.89 & 1.69\\
        ${E}_{b}^{c}$ (eV) & 0.095 & 0.113 & 0.107 & 0.107\\
        ${W}_{{ij}}$ (${{10}}^{{16}}{\rm s}^{-1}$) & 7.30 & 2.45 & 69.7 & 7.44\\
    \bottomrule
    \end{tabular}
\end{table}

\clearpage
\begin{table}[t]
    \centering
    \caption{{Calculated electron capture cross-section of $E'_\gamma$ and underlying parameters.}}
    \label{tab:my_label}
    \begin{tabular}{ccccc}
    \toprule
        {\bf Sample No.} & {\bf 1} & {\bf 2} & {\bf 3} & {\bf 4}\\
    \midrule
        ${\sigma}_{\gamma,e}$ (${{10}}^{-{14}}{\rm cm}^2$) & 3.20 & 4.54 & 2.56 & 0.245\\
        ${E}_{{ij}}$ (eV) & 2.20 & 2.04 & 2.09 & 2.23\\
        ${\lambda}_{{ij}}$ (eV) & 1.14 & 1.07 & 1.04 & 1.13\\
        ${E}_{b}^{c}$ (eV) & 0.246 & 0.220 & 0.265 & 0.268\\
        ${W}_{{ij}}$ (${{10}}^{{14}}{\rm s}^{-1}$) & 2.10 & 2.98 & 1.68 & 0.161\\
    \bottomrule
    \end{tabular}
\end{table}

\end{document}